\documentclass[reprint,showpacs,preprintnumbers,amsmath,amssymb,aps,prb,floatfix]{revtex4-1}

\usepackage{graphicx}
\usepackage{dcolumn}
\usepackage{bm}

\begin{document}

\title{Gate-induced magneto-oscillation phase anomalies in graphene bilayers}
\author{N.\ A.\ Goncharuk}
\author{L.\ Smr\v{c}ka}
\affiliation{Institute of Physics, Academy of Science of the Czech
Republic,\nolinebreak[5] v.v.i.,\\ 
Cukrovarnick\'{a} 10, 162 53 Prague 6, Czech Republic}

\date{\today}

\begin{abstract}
The magneto-oscillations in graphene bilayers are studied in the
vicinity of the $K$ and $K'$ points of the Brillouin zone within the
four-band continuum model based on the simplest
tight-binding approximation involving only the nearest neighbor
interactions. The model is employed to construct Landau plots for
a variety of carrier concentrations and bias strengths between the
graphene planes. The quantum-mechanical and quasiclassical approaches
are compared. We found that the quantum magneto-oscillations are only
asymptotically periodic and reach the frequencies predicted
quasiclassically for high indices of Landau levels. In unbiased
bilayers the phase of oscillations is equal to the phase of massive
fermions. Anomalous behavior of oscillation phases was found in
biased bilayers with broken inversion symmetry. 
The oscillation frequencies again tend to quasiclassically predicted ones,
which are the same for $K$ and $K'$, but the quantum approach yields the
gate-tunable corrections to oscillation phases, which differ in sign
for $K$ and $K'$.  These valley-dependent phase corrections give rise,
instead of a single quasiclassical series of oscillations, to two
series with the same frequency but shifted in phase.
\end{abstract}

\maketitle

\section{\label{intro}Introduction}
In solids subject to a magnetic field $B$, the energy spectrum of charge
carriers is quantized into Landau levels (LLs). The
magneto-oscillations (MOs) observed in the Shubnikov-de Haas and
de Haas-van Alphen effects reflect the oscillations of the
density of states (DOS) with the field intensity. The DOS reaches
maxima at magnetic fields, $B_n$, for which the LLs with the index, $n$,
cross the Fermi energy, $E_F$.

The Landau plot is a plot of inverse magnetic fields, $1/B_{n}$,
versus the LL index, $n$. It is a standard tool used to determine the
frequency and phase of MOs, and the related important
characteristics of the investigated systems.

The construction of the Landau plot is based on the Onsager-Lifshitz
quasiclassical quantization rule, \cite{Onsager,Lifshitz}
\begin{equation} 
A(E_F)= \frac{2\pi|e|B}{\hbar}\left(n+\gamma\right), \
\label{onsager}
\end{equation}
where $A(E_F)$ is an area of the extremal cross-section of the Fermi
surface (FS) cut by the plane perpendicular to the magnetic field
direction, $e$ is the electron charge and $\gamma$ is a constant which
describes the phase of MOs. It follows from Eq.~(\ref{onsager}) that MOs
of DOS are periodic in $1/B$ and their frequency $F$ is related to
$A(E_F)$ by
\begin{equation} 
F = \frac{\hbar A(E_F)}{2\pi|e|}.
\label{quasi}
\end{equation}
The Onsager-Lifshitz quantization rule has been originally designed for
three dimensional metals, where the validity of the
quasiclassical approximation is guaranteed by a large number of LLs
bellow $E_F$ in accessible magnetic fields. However, the method is also
widely used when two-dimensional (2D) systems are investigated. Here, 
the importance of $F$ is stressed out by the fact that the
carrier concentration is proportional to the area surrounded by the Fermi
contour.

In general, the rule should not be applicable to 2D systems. Subject to
strong magnetic fields, the quantum limit with only one LL below $E_F$ 
can be easily reached. But in the majority of such systems, 
the periodicity of MOs is preserved due to the simple parabolic
(Schr\"{o}dinger--like) energy spectra of the 2D electron layers in the
semiconductor structures, which yields the LL energies proportional to
$B$.

In 2004 a single sheet of graphene was separated from bulk graphite by
micromechanical cleavage. \cite{geim05} It was confirmed
experimentally that electrons in graphene obey a linear energy
dependence on the wave-vector $\vec{k}$, as predicted many years ago
by the band structure calculation. \cite{wallace} Both electron and
hole charge carriers  behave like massless relativistic
particles -- Dirac fermions (DFs), and there is no gap between the
valence and conduction bands. The electron and hole Dirac cones touch
at a neutrality point.

Subject to a magnetic field $B$, the DFs form LLs
with energies proportional to $\sqrt{B}$. In the seminal papers
\cite{geim04, geim05,kim05} the Shubnikov-de Haas MOs in graphene were found
periodic in $1/B$, similarly as in the 2D gas of Schr\"{o}dinger
fermions (SFs) with the parabolic energy spectra, but with the phase
shifted by $\pi$. The shift, which was clearly demonstrated by the
Landau plot of magneto-resistance oscillations, is due to the
existence of the zero-energy LL in the linear Dirac spectrum, shared
by electrons and holes.  Note that $\gamma=1/2$ for SFs, and
$\gamma=0$ for DFs.

In addition to a single layer graphene, also a few layer graphene
samples can be prepared.  Among them a bilayer graphene (BLG), in which two
carbon layers are placed on top of each other with a standard Bernal
stacking, is of particular interest. Probably the most remarkable
feature of this structure is the possibility to open a gap between the
valence and conduction bands through the application of an external
field or by chemical doping.~\cite{McCann,Ohta,ECastro-Geim}. This
phenomenon is closely related to the gate-induced breaking of the
inversion symmetry of the crystal.\cite{xiao,mucha,nakamura,zhao}

Note also that the application of the gate voltage is a necessary
condition for the experimental observation of MOs in BLG. 
Without a gate voltage, the sample is neutral, the Fermi
energy is located in the neutrality point, and no free charge carriers
should be present in perfect samples.

There are two ways of how to apply the gate voltage. If the external
voltage is applied symmetrically from both sides of a sample, just $E_F$
and the concentration of carriers are varied, and no gap is opened. The
tunable gap appears in the presence of external electric field resulting
from the asymmetrically applied gate voltage.

Let us point out that the charge carriers in BLG are
neither SFs nor DFs, and therefore it is of interest to construct the
corresponding Landau plots to see how far the bilayer energy spectra
from these two simplest possibilities are.

This task is simplified by the fact that the electrochemical potential
(i.e., also $E_F$) is kept constant during magnetic field sweeps in
gated samples. According to Ref.~\onlinecite{mosser,imura}
carrier density oscillations are compensated by gate current
oscillations in the case of fixed $E_F$. Note that in bulk samples,
where the charge neutrality must be preserved, the carrier concentration
is considered to be fixed.

To construct the Landau plot, we will first calculate the quasiclassical
frequencies of MOs in BLG, based on their
zero-magnetic-field electronic structure.

Later on we will compare these quasiclassical frequencies with results
of the quantum-mechanical calculation of the electronic structure of
BLG subject to a perpendicular magnetic field.
%
\section{\label{zero}Zero-field electronic structure}
The electronic structure of BLG can be described by the
simple tight-binding model involving only the nearest neighbor
interactions.\cite{Pereira,McCann1,McCann2,Nilsson,ECastro,Koshino,
CastroNeto-Geim}

A single layer honeycomb lattice, with two atoms per unit cell, results
from two superimposed triangular lattices labeled A and B.  The unit
cell is defined by the lattice vectors $\vec{a}_1$ and $\vec{a}_2$, making
the angle 60$^\circ$, the lattice constant $a$ is equal to 2.46 \AA.
%
\begin{figure}[htb]
\includegraphics[width=0.8\linewidth]{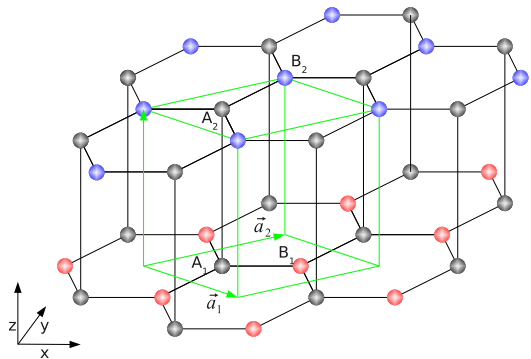}
\caption{\label{fig1} (Color online) Lattice structure of a graphene 
bilayer. The unit cell is a green parallelepiped.}
\end{figure}
%
The bilayer is formed by two graphene sheets, 1 and 2, arranged in the
Bernal stacking. The distance between layers is 3.37 \AA. Thus the
unit cell of a bilayer has four atoms, its lattice structure is 
sketched in Fig.~\ref{fig1}.

In addition to the intralayer parameter $\gamma_0$ and the interlayer
parameter $t$, the corresponding Hamiltonian depends on the potential energy
difference between the two layers, which we denote $2u$.  The
parameter $\gamma_0 \approx 3.1$ eV yields the Fermi velocity $v_F
\approx 1.0\times 10^6 $ m/s, defined by $\hbar v_F =
\gamma_0\sqrt{3}a/2$. We further consider that $t \approx 0.39$ eV,
and the energy $2u$ varies between $0$ and $250$ meV.\cite{zhang}
While $\gamma_0$ and $t$ are fixed by nature, we assume that $u$ and
$E_F$ are the adjustable parameters.

If we employ the continuum approximation, \cite{wallace} the Hamiltonian
$H$ in the vicinity of the $K$ point can be written as
\begin{equation}
H=
\left(\begin{array}{cccc}
H_{B_2B_2} & H_{B_2A_2} & H_{B_2A_1} & H_{B_2B_1}\\
\\
H_{A_2B_2} & H_{A_2A_2} & H_{A_2A_1} & H_{A_2B_1}\\
\\
H_{A_1B_2} & H_{A_1A_2} & H_{A_1A_1} & H_{A_1B_1}\\
\\
H_{B_1B_2} & H_{B_1A_2} & H_{B_1A_1} & H_{B_1B_1}\\
\end{array}\right),
\label{Hamilton}
\end{equation} 
where the matrix elements of the first layer are given by
\begin{eqnarray}
H_{A_1A_1}& =&H_{B_1B_1}= -u ,\nonumber\\ 
H_{A_1B_1}&=&H^*_{B_1A_1}=\hbar v_F(k_x-i k_y) . \nonumber
\end{eqnarray}
Similarly,
the matrix elements corresponding to the second layer read
\begin{eqnarray}
H_{A_2A_2}&=&H_{B_2B_2}= u , \nonumber\\
H_{A_2B_2}&=&H^*_{B_2A_2}= \hbar v_F(k_x+i k_y) . \nonumber
\end{eqnarray}
There
are only two nonzero interlayer matrix elements
\begin{equation}
H_{A_1A_2} =H_{A_2A_1}= t . \nonumber
\end{equation}

The Hamiltonian $H'$ in the vicinity of the $K'$ point has a similar
structure, the matrix elements of $H'$ are complex conjugates of the 
matrix elements of $H$.
%
\begin{figure}[htb]
\includegraphics[width=0.8\linewidth]{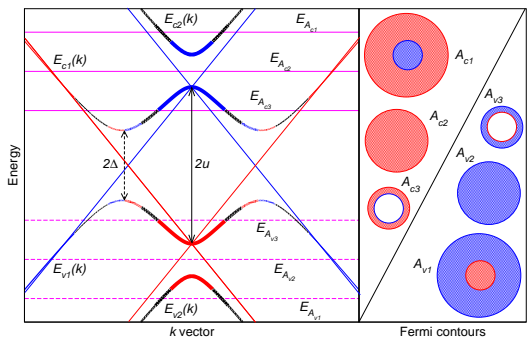}
\caption{\label{fig2}(Color online) The ,,mexican hat'' shape of the
valence and conduction bands of a biased bilayer. The blue and red
colors correspond to higher probability of finding charge carriers in
the layers 1 and 2, respectively. Three groups of the Fermi contour are
possible depending on the value of $E_F$: the double circles (A1), the
circles (A2), and the Fermi rings (A3).}
\end{figure} 

The above Hamiltonians can be diagonalized
analytically.\cite{Pereira,Nilsson,ECastro,ECastro1,CastroNeto-Geim}
The zero-field energy branches of the conduction band, $E_{c1}(k)$ and
$E_{c2}(k)$, and the valence band, $E_{v1}(k)$ and $E_{v2}(k)$, of a
bilayer result from hybridization of Fermi cones of layers 1 and 2,
mediated by the interlayer matrix element $t$. Note that 
$E_{v1}(k)=-E_{c1}(k)$ and $E_{v2}(k)=-E_{c2}(k)$ and that the valley
degeneracy is preserved, i.e., we get the same bands in
valleys $K$ and $K'$.

For $u=0$ two Fermi cones are replaced by four bonding and antibonding
hyperbolic bands. The bonding valence and conduction bands,
$E_{v1}(k)$ and $E_{c1}(k)$, touch at $k=0$, the separation between
bands of a bonding--antibonding pair is equal to $t$ on the energy
scale.
 
When the interlayer voltage is applied, the Fermi cones of two layers
are shifted along the energy axis, and the separation of the neutrality
points becomes equal to $2u$. The hybridization due to the interlayer parameter
$t$ is strongest near the cone cross-points. The resulting four bands are
shown in Fig.~\ref{fig2}. It turns out that for any finite $u$ a gap is
open between the topmost valence band $E_{v1}(k)$ and the bottom
conduction band $E_{c1}(k)$. The conduction band acquires a ,,mexican
hat'' shape with energy minima at nonzero $k$ and a local maximum at
$k=0$.  We can write
\begin{eqnarray}
E_{c1}^{\,\,\text{max}}(0)&=&u,  \nonumber\\
E_{c2}^{\,\,\text{min}}(0)&=&\sqrt{u^2+t^2}, \nonumber\\
E_{c1}^{\,\,\text{min}}(k)&=&\Delta=ut/\sqrt{4u^2 +t^2}.
\label{minmax}
\end{eqnarray}
Note that for large $k$ the band $E_{c1}(k)$ describes electrons
localized mostly in the layer 1. Near the local maximum at $k=0$ the
holes in the layer 2 prevail. Similar conclusions can be drawn for the
topmost valence band $E_{v1}(k)$.

As mentioned in Introduction, the quasiclassical frequencies of the
bilayer, $F_1$ and $F_2$, are proportional to areas surrounded by the
Fermi circles, which depend, for a given $u$, on the Fermi energy
value $E_F$. Three different possibilities are depicted in
Fig.~\ref{fig2} for the case of conduction/valence bands. (For the
valence bands $E_F$ should be replaced by $-E_F$.)

The analytic expressions for the quasiclassical frequencies $F_1$ and
$F_2$ read
\begin{equation}
F_{1(2)}  = \frac{2\hbar}{3|e|a^2\gamma_0^2}\left[E_F^2+u^2 
\pm \sqrt{(E_F^2-u^2)t^2 + 4 E_F^2u^2}\right],
\label{B12u}
\end{equation}
the frequencies $F_{1(2)}$ are even functions of variables $E_F$ and
$u$.
The frequency $F_2$ is equal to zero at the local maximum
$E_{c1}^{\,\,\text{max}}(0)$, and at the minimum
$E_{c2}^{\,\,\text{min}}(0)$.  For a finite $u$, the frequency $F_2$
approaches $F_1$ at $E_{c1}^{\,\,\text{min}}(k)$.

Three forms of the Fermi contour are possible depending on the value
of $E_F$.  First, the large $E_F$ cuts both conduction bands and 
$F_1 > F_2 > 0$. The frequency $F_1$ corresponds to electron orbits
localized mainly in the layer 1, the frequency $F_2$ corresponds to
hole orbits localized mainly in the layer 2.  Second, only the band
$E_{c1}$ is cut by $E_F$. Then $F_1 > 0$ and $F_2 < 0$. In this case
$F_2$ is just a parameter and does not have the meaning of a true
frequency.  At last, the $E_F$ cuts the bottom conduction band
$E_{c1}(k)$ twice, if it is less than a local energy maximum, 
$E_{c1}^{\,\,\text{max}}(0)$. 
Then again $F_1 > F_2 > 0$. In that case $F_1$ is the
frequency of an electron orbit in the layer 1 while $F_2$ belong to a
hole orbit in the layer 2. Close to the local minima the difference
between electrons and holes is smeared and charge carriers are
present in both layers as indicated by the change of line colors in
Fig.~\ref{fig2}.

For the special case of $u=0$, Eq.~(\ref{B12u}) reduces to
\begin{equation}
F_{1(2)}  = \frac{2\hbar}{3|e|a^2\gamma_0^2}
\left(E_F \pm t\right)E_F. 
\label{B12}
\end{equation}
Then the gap between the valence and conduction bands as well
as the local maximum $E_{c1}^{\,\,\text{max}}(0)$ all disappear.

The quasiclassical phases of MOs are not accesible via the
Onsager-Lifshitz quantization rule, Eqs.~(\ref{onsager}) and
(\ref{quasi}).  To find the energy spectra beyond the quasiclassical
approximation, we need to diagonalize the magnetic Hamiltonians $H$ and
$H'$.
\section{\label{nonzero}Magnetic field effects}
The magnetic Hamiltonians can be obtained from the zero-field
Hamiltonians by modification of matrix elements $H_{A_1B_1}$,
$H_{B_1A_1}$, $H_{A_2B_2}$ and
$H_{B_2A_2}$.\cite{MacClure60,Inoue,Pereira} 
The matrix elements of the
magnetic Hamiltonian in the vicinity of the $K$ point are
\begin{eqnarray}
H_{A_1B_1}& =&H^*_{B_1A_1}=\sqrt{2|e|\hbar v_F^2 B\,n},\nonumber\\
H_{A_2B_2}& =&H^*_{B_2A_2}=\sqrt{2|e|\hbar v_F^2 B\,(n+1)}. \nonumber
\end{eqnarray}
%
The other matrix elements remain the same as in the zero-field
Hamiltonian. Near the $K'$ point, 
\begin{eqnarray}
H'_{A_1B_1}& =&H'^*_{B_1A_1}=\sqrt{2|e|\hbar v_F^2 B\,(n+1)},\nonumber\\
H'_{A_2B_2}& =&H'^*_{B_2A_2}=\sqrt{2|e|\hbar v_F^2 B\,n}. \nonumber
\end{eqnarray}
%

We need not diagonalize these Hamiltonians to construct the Landau
plot. If we look for magnetic fields $B_n$ at which the LLs cross
$E_F$, it is enough to find the poles of the resolvent
$G(z)=(z-H)^{-1}$, as it defines the density of states $g(E_F)$
through
\begin{equation} 
g(E_F) \propto -\frac{1}{\pi}\, {\mathcal Im}\, {\text Tr}\,G(E_F+i0 ).
\label{res}
\end{equation}
The easiest way to find the poles is to solve the corresponding
secular equation for $B_n$ assuming the fixed $E_F$.

We start with the simplest case of the unbiased BLG ($u=0$). Then
the secular equations can be given a very convenient form, utilizing
the quasiclassical frequencies of MOs, presented in the previous
paragraph, Eq.~(\ref{B12}),
\begin{equation}
B^2 n(n+1) -B\left(n+\frac{1}{2}\right)(F_1+F_2)+F_1F_2 =0.
\label{sec_eq}
\end{equation}
The Hamiltonians $H$ and $H'$ yield identical equations for valleys 
$K$ and $K'$.

While the secular polynomial is quartic in energy it is only quadratic
in $B$. Therefore, to construct the Landau plot it is enough to solve
the quadratic equation to find $B_n$ in terms of fixed $E=E_F$,.

The quasiclassical phase $\gamma$ can be easily obtained from
Eq.~(\ref{sec_eq}). For a large number of LLs below $E_F$ one may
assume that $n(n+1)\rightarrow (n+1/2)^2$, and then Eq.~(\ref{sec_eq})
can be written in the form
\begin{equation}
B^2\left(n+\frac{1}{2}\right)^2 -B\left(n+\frac{1}{2}\right)(F_1+F_2)+F_1F_2 =0.
\label{sec_eq_qcl}
\end{equation}
From here we obtain the asymptotic quasiclassical Landau plots 
\begin{equation}
\frac{F_{1(2)}}{B_n} =n +\frac{1}{2},
\end{equation}
i.e., we found that the phases of MOs correspond to SFs with $\gamma
=1/2$, in agreement with quasiclassical treatments of systems with
inversion symmetry. Note that $F_2$ is positive only in the
rather unrealistic case $|E_F|>t$.

To get Landau plots for an arbitrary $n$ we can express the solution of
Eq.~(\ref{sec_eq}) as
\begin{equation}
\frac{2F_1F_2}{F_1+F_2}\frac{1}{B_n}= n+\frac{1}{2}\mp 
  \sqrt{\left(n+\frac{1}{2}\right)^2 -n(n+1)\frac{4F_1F_2}{(F_1+F_2)^2}}
\end{equation}
or, if we define dimensionless $\delta$ by
\begin{equation}
\delta = \left(\frac{F_1-F_2}{F_1+F_2}\right)^2,
\end{equation}
we can write (see also Ref.~\onlinecite{Smrcka}) 
\begin{equation}
\frac{F_{1(2)}}{B_n} = 
\frac{n+\frac{1}{2}\mp \sqrt{\frac{1}{4}+n(n+1)\delta}} 
{1\mp\sqrt{\delta}}.
\label{freq}
\end{equation}
Here the negative sign in the numerator/denominator corresponds to the
frequency $F_1$ in the quasiclassical limit, and the positive sign to
the quasiclassical frequency $F_2$. It is obvious that for $ \delta \ne
0$ the MOs are not periodic in $1/B$.

The case of the biased BLG ($u\neq 0$) must be treated separately,
as the presence of the electric field perpendicular to layer planes
breaks the inversion symmetry and lifts the valley
degeneracy.\cite{mucha,nakamura}

The secular equations can be given again a form quadratic in $B$, but
the coefficients do not depend exclusively on the quasiclassical
frequencies as in Eq.~(\ref{sec_eq}). We can write
\begin{equation}
B^2 n(n+1)-B\left[(n+\frac{1}{2})(F_1+F_2)+
F_0\right]+F_1F_2 =0
\label{sec-eq-bi}
\end{equation}
for the Hamiltonian $H$ in the vicinity of $K$.
In comparison with  Eq.~(\ref{sec_eq})  there is an extra  term 
\begin{equation}
F_0 = \frac{4\hbar}{3|e|a^2\gamma_0^2} E_F u.
\end{equation}
In the vicinity of $K'$ we obtain a very similar equation from 
the Hamiltonian $H'$, the only difference is that $F_0$ is replaced by
$-F_0$.  The extra term, $F_0$, is the reason of the valley asymmetry. It is
obvious that Eq.~(\ref{sec-eq-bi}) gives two different series of
solutions, $B_n$, for positive and negative $F_0$.

The quasiclassical frequencies $F_1$ and $F_2$ are even functions of
$E_F$ and $u$. It means that there are the same frequencies not only for
$K$ and $K'$, but also for the electrons and holes with energies $E_F$
and $-E_F$, respectively. 
Note also that $F_1$ and $F_2$ do not depend on the sign of
$u$. On the other hand, $F_0$ is an odd function of $E_F$ and $u$. Thus
$F_0$ breaks the $K$ -- $K'$ symmetry, and also the symmetry between the
electron and hole oscillations with the same quasiclassical
frequencies. The change of sign of $u$ also reverts the roles of $K$ and
$K'$ valleys, i.e., what is valid for $K$ with $u>0$ is valid for $K'$
with $u<0$.

Also Eq.~(\ref{sec-eq-bi}) can be rewritten to an equation similar
to Eq.~(\ref{freq}), but with an additional dimensionless parameter
$\lambda$, which depends on $F_0$,
\begin{equation}
\lambda = \frac{F_0}{F_1+F_2}.
\end{equation}
Then the analytic solution reads
\begin{equation}
\frac{F_{1(2)}}{B_n} = 
\frac{n+\frac{1}{2}+\lambda\mp\sqrt{(n+\frac{1}{2}+\lambda)^2
+n(n+1)(1-\delta)}} 
{1\pm\sqrt{\delta}}.
\label{freq_u}
\end{equation}
This equation reduces to Eq.~(\ref{freq}) for $\lambda = 0$.
 
Now it is a more difficult task to find an asymptotic expression for the
oscillation phase than in the previous case $u=0$. If we solve 
Eq.~(\ref{sec-eq-bi}) for $n+1/2$ we get
\begin{equation}
n+\frac{1}{2}=\frac{F_1+F_2\pm\sqrt{(F_1-F_2)^2+4BF_0+B^2}}
{2B},
\end{equation}
which for  $B$ approaching zero yields
\begin{equation}
\frac{F_{1(2)}}{B} = n+\frac{1}{2}\mp \xi.
\label{extra}
\end{equation}
Here $\xi$ is a gate-tunable correction to the oscillation phase, given by
\begin{equation}
\xi = \frac{F_0}{F_1-F_2 } = \frac{E_F u}{\sqrt{(E_F^2-u^2)t^2 +4E_F^2u^2}}.
\label{extra1}
\end{equation}
This correction differs in sign for $K$ and $K'$ and also differs for
electrons and holes from the same valley with the same absolute value
of energy.
%
\section{\label{results}Results and discussion}
\begin{figure}[b]
\includegraphics[width=0.8\linewidth]{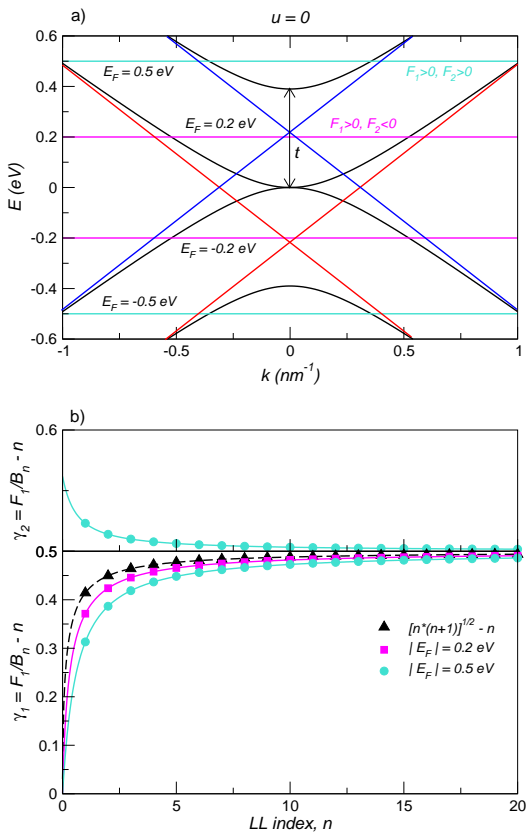}
\caption{\label{fig3} (Color online) 
(a) The electronic bands of the unbiased graphene bilayer ($u=0$). The
horizontal lines denote the Fermi energies which cross the electron
and hole dispersion curves.  (b) The ,,phases''
$\gamma_{1(2)}(E_F)=F_{1(2)}/B-n$, for $E_F$ depicted in a), plotted as
functions of the Landau level index~$n$.}
\end{figure}
In the unbiased BLG the energy $u$ is equal to zero and the
parameter $\delta$, which appears in Eq.~(\ref{freq}), has a
particularly simple form
\begin{equation}
\delta = \frac{t^2}{E_F^2}.
\end{equation}
%

\begin{figure}[b]
\includegraphics[width=0.8\linewidth]{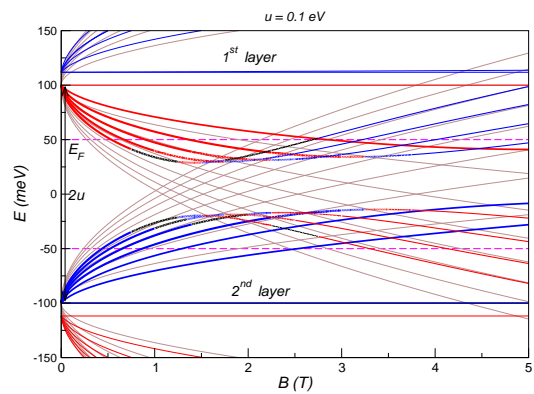}
\caption{\label{fig4} (Color online) The electron and hole Landau
levels (in the $K'$ valley) 
of two layers are mixed by the interlayer interaction, $t$. For
the energy range corresponding to Fermi rings in zero magnetic filed
(see $A_3$ in Fig.~\ref{fig2}),
$E_F$ cuts the Landau levels twice. This is the reason for the
anomalous phase in the quasiclassical limit $B \rightarrow 0$.}
\end{figure}
For small Fermi energies only the bottom branch $E_{c1}(k)$ of the
conduction subband is cut by $E_F$ and only the frequency $F_1$ is
defined. For $E_F$ approaching zero, the parameter $\delta$
diverges. This implies that for energies close to the band bottom
Eq.~(\ref{freq}) can be written as
\begin{equation}
\frac{F_1}{B} = \sqrt{n(n+1)},
\end{equation}
the form found for the extremal electron and hole orbits in
graphite,\cite{Smrcka} which clearly indicates the aperiodicity of
oscillations.

The Fermi energies greater than $t$ are rather
unrealistic. Nevertheless we can consider this hypothetical case in
our theoretical treatment. We can write, for $E_F =t$ and $\delta =1$,
Eq.~(\ref{freq})  as follows
\begin{eqnarray}
\frac{F_1}{B}& = &\frac{n(n+1)}{ n+\frac{1}{2}},\\
\frac{F_1}{B}& = & n+\frac{1}{2}. \nonumber
\end{eqnarray}
The Landau plots calculated for two selected values of $E_F$, $E_F<t$
and $E_F>t$, which cross the dispersion curves are presented in
Fig.~\ref{fig3}. The Landau plots are the same for $E_F$ and $-E_F$
due to the inversion symmetry conservation.  One can observe that in the
unbiased bilayer the phases of MOs, corresponding to both frequencies
$F_1$ and $F_2$, approach the phase of massive fermions,$\gamma=1/2$,
for higher quantum numbers of LLs.

In BLG, an applied electric field leads to asymmetry
between $K$ and $K'$ valleys that gives rise to nontrivial oscillation
phenomena in magnetic fields. To illuminate an anomalous
behavior of oscillations, we plotted in Fig.~\ref{fig4} the field
dependence of LLs in BLG.

In a single layer graphene the LL fans of electrons and holes start at
the zero-field neutrality point.  The neutrality points of two
independent layers are shifted by $2u$ and the LLs of holes from the layer
1 cross the LLs of electrons from the layer 2, as shown in Fig.~\ref{fig4}
by thin brown lines.

In BLG the shape of LL spectrum results from
hybridization of LL spectra of layers 1 and 2.  Due to the interlayer
interaction, represented by the matrix element $t$, we have four fans
of LLs which start at zero-field energies $E_{v2}(0)$, $E_{v1}(0)$,
$E_{c1}(0)$ and $E_{c2}(0)$.

The hole levels from the layer 1 and the electron levels from the
layer 2 avoid to cross, and the low-field hole LLs smoothly turn into
the electron LLs as $B$ increases. This is indicated in
Fig.~\ref{fig4} by the change of LL color from red to blue. The LLs
from a fan starting at zero-field energy $E_{c1}(0)$ have minima in
their field dependence and, therefore, can be cut twice by a single
$E_F$. Moreover, the minima are not the same for all levels and,
consequently, not all levels are cut by a single $E_F$.

This is reflected in the quasiclassical approach as the gate-dependent
correction to the MO phase, $\xi$, which is related to the energy
difference $2u$ between two layers. Note that in the region of energies
corresponding to the Fermi rings the expression~(\ref{extra1}) diverges
at $E_{c1}^{\,\,\text{min}}(k)$ and is equal to $1/2$ for $E_F =
E_{c1}^{\,\,\text{max}}(0)$.  As the many body effects can play a role
in this low concentration range, the above one-electron picture is probably
oversimplified.
\begin{figure}[tb]
\includegraphics[width=0.8\linewidth]{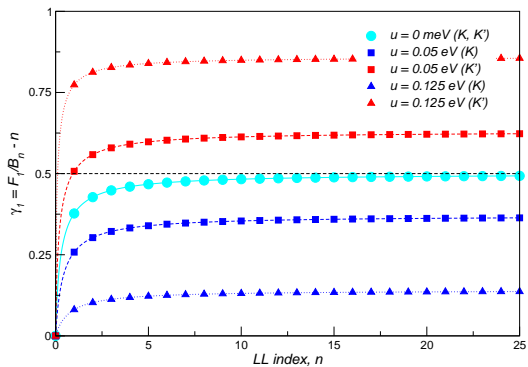}
\caption{\label{fig5} (Color online) The ,,phase'' 
$\gamma_{1}=F_{1}/B-n$ calculated for the fixed quasiclassical
 frequency $F_1=70$~T and various $u$, for the
 electron $K$ and $K'$ valleys as a function of the LL index, $n$.}
\end{figure}
\begin{figure}[b]
\includegraphics[width=0.8\linewidth]{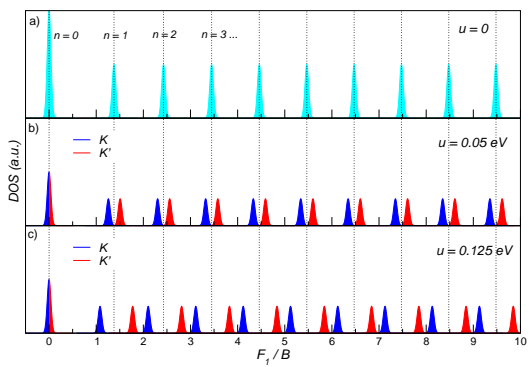}
\caption{\label{fig6} (Color online) The DOS of the
unbiased (a) and biased (b, c) BLG versus dimensionless
value of the Landau plot, $F_1/B$, with the fixed quasiclassical 
frequency $F_1=70$~T. The frequency $F_1$ corresponds to the
situation when $F_1>0, F_2<0$, i.e.,  only the lowest
conduction/valence energy band is cut by $E_F$. 
In (b, c) the blue peaks show DOS calculated for the $K$
valley, whereas the red ones are related to the $K'$ valley.}
\label{Lala}
\end{figure}
\begin{figure}[hb]
\includegraphics[width=0.75\linewidth]{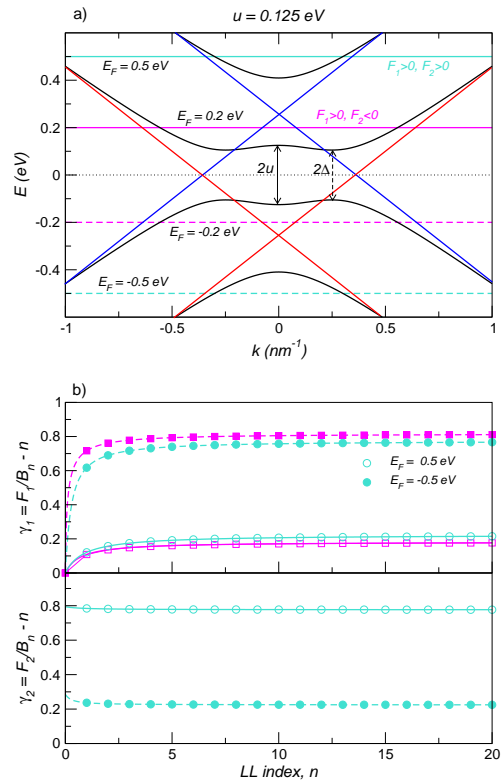}
\caption{\label{fig7}(Color online)
(a) The electron/hole bands of the biased graphene bilayer with the gap
 $2u=0.25$~eV at $k=0$. The horizontal
lines denote the Fermi energies which cross the electron (solid lines) 
and hole (dashed lines) dispersion curves. 
(b) The ,,phases'' $\gamma_{1(2)}(E_F)=F_{1(2)}/B-n$
calculated for $E_F$ depicted in (a) plotted as functions of the LL
index, $n$.} 
\end{figure}
\begin{figure}[ht]
\includegraphics[width=0.8\linewidth]{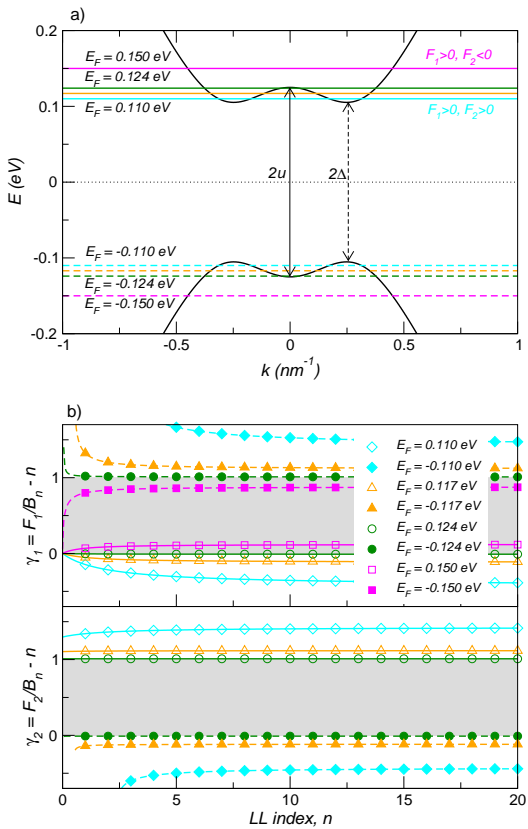}
\caption{\label{fig8}(Color online)
The same as in Fig.~\ref{fig7}, only for the lowest electron/hole bands
 and $\Delta<|E_F|<u$.} 
\end{figure}

We start our discussion with the simplest case,
$E_{c1/v1}(0)<|E_F|<E_{c2/v2}(0)$, when only one conduction/valence
energy band is cut by $E_F$ (see Fig.~\ref{fig2}).  The
single quasiclassical frequency $F_1$ corresponds to a single Fermi
area, which is the same for $K$ and $K'$.  

According to Eq.~(\ref{extra}), the electron peaks in DOS are shifted by
$\xi$ to the left in the $K$ valley, whereas the peaks in the $K'$
valley are shifted by $\xi$ to the right. The  shift magnitude
ranges from zero to $1/2$ depending on the energy difference between two
layers.  In Fig.\ref{fig5} the ,,phase'' $\gamma_1$ is plotted as the
function of $n$ for three different cases with $u$ equal to 0, 0.05 and
0.125 eV. The Fermi energies are chosen to keep the same Fermi area
(and the same fixed $F_1$) in all three systems. Only for $u=0$ the curves
are identical for $K$ and $K'$, for $u\neq 0$ the curves are
substantially different.

In Fig.~\ref{fig6} the shifts of peaks in the above three cases are shown
explicitly. There is a single series of oscillations for the unbiased
bilayer, as the valley degeneracy is preserved in a system with the
inversion symmetry.

The effect of the gate-tunable valley splitting originates in two
series of oscillations which differ for the different $u$.  Let us
emphasize that all series of oscillations have the same
quasiclassical frequency $F_1$, but the quasiclassical phases
depend on the choice of $u$ and on the choice of the valley index.

We complete our discussion with cases when the Fermi energy cuts the
conduction/valence bands twice. The Landau plots of the biased
bilayer with $u=0.125$~eV, which is probably the highest 
experimentally accessible value,~\cite{zhang} 
calculated for four selected Fermi energies, two in the
conduction band, and two in the valence band, are presented in
Fig.~\ref{fig7}. In accordance with types of the Fermi contours in
Fig.~\ref{fig2}, the first and second cases are depicted.

The situation is more complicated in the region of energies,
$\Delta<|E_F|<u$, for which $E_F$ cuts the lowest subband
$E_{c1/v1}(k)$ twice, which is characteristic for the third type of
the Fermi contours, as shown in Fig.~\ref{fig2}. The bottom of
$E_{c1}(k)$ is at $\approx 0.105$~eV.  The parameter $\xi$ is far from
the values expected for the phase of quasiclassical oscillations, it
reduces/grows heavily when $E_F$ approaches the bottom of
$E_{c1}(k)$/$E_{v1}(k)$.  For $E_F=u$, $\xi$ becomes closer to $-1/2$
for $E_F$ in the conduction band and $1/2$ for $E_F$ in the valence
band.\\

%
\section{\label{cocl}Conclusions}
Using a four-band continuum model, we calculated analytically the
Landau plots in biased and unbiased BLG subject to
external perpendicular magnetic fields.

It turns out that the magneto-oscillations are only asymptotically
periodic, and that in the unbiased bilayers their phase is equal to
the phase of massive fermions. The convergence to the quasiclassical
limit is slow, and depends strongly on the the value of $E_F$. The
convergence is slower for higher values of $E_F$.

Anomalous behavior of oscillation phases was found in biased bilayers
with broken inversion symmetry.  The oscillation frequencies again
tend to quasiclassically predicted ones, which are the same for $K$
and $K'$, but the quantum approach yields the gate-tunable corrections
to oscillation phases, which differ in sign for $K$ and $K'$.  These
valley-dependent phase corrections give rise, instead of a single
quasiclassical series of oscillations, to two series with the same
frequency but shifted in phase.

We also found that for $E_F$ in the region of energies corresponding
to the Fermi rings in the quasiclassical approach, only a limited
number of LLs can cut the Fermi energy and thus a limited number of
magneto-oscillations can be achieved. Moreover, their quasiclassical
phases reach very large values. As the many body effects can play a
role in the corresponding concentration range, the above one-electron
picture is probably oversimplified.
\begin{acknowledgments}
The authors acknowledge the support of the Academy of Sciences of the
Czech Republic project KAN400100652, the Ministry of Education of the
Czech Republic project LC510, and the PHC Barrande project 19535NF and MEB
020928.
\end{acknowledgments}

\bibliography{text.bib}

\end{document}